\documentstyle[aps,preprint,epsf,prb]{revtex}
\tightenlines

\begin{document}

\title{A case study of stratus cloud base height multifractal fluctuations}

\author{ K. Ivanova,$^1$ H.N. Shirer,$^1$ E.E. Clothiaux,$^1$ N. Kitova,$^{2}$
M.A. Mikhalev,$^2$ T.P. Ackerman,$^3$ and M. Ausloos$^4$} \address{
$^{1}$Department of Meteorology, Pennsylvania State University, 
University Park,
PA 16802, USA \\ $^{2}$Institute of Electronics, Bulgarian Academy of Sciences,
72 Tzarigradsko chauss\'ee, Sofia 1784, Bulgaria\\ $^3$Pacific 
Northwest National
Laboratory, Richland, WA 99352, USA\\ $^4$SUPRAS \& GRASP, B5, Sart Tilman,
B-4000 Li$\grave e$ge, Euroland }



\maketitle

\begin{abstract} The complex structure of a typical stratus cloud 
base height (or
profile) time series is analyzed with respect to the variability of its
fluctuations and their correlations at all experimentally observed temporal
scales. Due to the underlying processes that create these time series, they are
expected to have multiscaling properties. For obtaining reliable measures of
these scaling properties, different methods of statistical analysis are used
herein : power spectral density, detrended fluctuation analysis, and 
multifractal
analysis. This broad set of diagnostic techniques is applied to a 
typical stratus
cloud base height (CBH) data set; data were obtained from the Southern Great
Plains site of the Atmospheric Radiation Measurement Program of the 
Department of
Energy from a Belfort Laser Ceilometer. First, we demonstrate that 
this CBH time
series is a nonstationary signal with stationary increments. Further, 
two scaling
regimes are found, although the characteristic laws are quite similar 
ones. Next,
the multi-affine scaling properties are confirmed. The scaling 
properties of the
cloud base height profile of such a continental stratus are found to be similar
to those of the marine cloud base height profiles studied by us 
previously. Some
physical interpretation in terms of anomalous diffusion (or fractional random
walk) is given for the continental case.\end{abstract}

\newpage

\section{Introduction}

The state of the atmosphere is governed by the classical laws of 
fluid motion and
exhibits a great deal of correlation at various spatial and temporal scales.
Knowing these space- and time-scale-dependent correlations is crucial, for
example, in order to understand the short and long term trends in climate. In
particular, clouds play an important role in the atmospheric energy budget. In
order to model and predict climate successfully, we must be able to 
both describe
the effects of clouds in the current climate and predict the complex chain of
events that might modify the distributions and properties of clouds 
in an altered
climate. Moreover, in order to improve the parameterization of clouds 
in climate
models, it is important to understand the cloud properties and their changes
within the cloud.

Modeling the impact of clouds is difficult because of their complex shapes,
spatial distributions and varying particle size distributions and because of
their differing effects on weather and climate. Clouds can reflect incoming
sunlight, and so contribute to cooling, but they also absorb infrared radiation
leaving the earth, and so contribute to warming. High cirrus clouds, 
for example,
may have a nonnegligible impact on atmospheric warming. Low-lying 
stratus clouds,
which are frequently found over oceans, can contribute to warming as
well.\cite{andrews}

A variety of physical processes takes place in the atmospheric 
boundary layer. At
time scales of less than one day, significant fluxes of heat, water vapor and
momentum occur due to entrainment, radiative transfer, and/or
turbulence.\cite{andrews,garratt} The turbulent character of the motion in the
atmospheric boundary layer (ABL) is one of its most important features. The
turbulence\cite{frisch} can be caused by a variety of processes, among them
thermal convection and mechanical generation by wind
shear.\cite{garratt,panofsky,dried} This complexity of physical processes and
interactions between them creates a variety of atmospheric responses. In
particular, in a cloudy ABL, the radiative fluxes produce local sources of
heating or cooling within the mixed layer and therefore can greatly 
influence its
turbulent structure and dynamics. Moreover, variations in the 
turbulent structure
and dynamics of the clouds cause subsequent changes in the cloud boundaries,
especially in the height of cloud base. These variations lead to a complex
structure for the cloud base height data (time) series. To analyze different
aspects of the variability of its fluctuations and correlations at all
experimentally observed temporal and spatial scales, one needs to 
apply a variety
of diverse techniques of statistical analysis to the retrieved time 
series data.
In what follows, we briefly present a few of these methods: a traditional
technique - power spectral density, and rather new techniques, like a detrended
fluctuation analysis, and multifractal analysis and then apply them 
to cloud base
height data.

\section{Scaling}

\subsection{Power spectral density}

The power spectral density $S(f)$ of a signal $y(t)$ is obtained as a Fourier
transform (FT) of the signal.\cite{FT,falconer} The power spectrum provides
information on the amplitude of the predominant frequencies present in the time
series. This information allows one to identify periodic, multi-periodic,
quasiperiodic or nonperiodic signals. Usually the logarithmic power 
spectrum plot
is used to better distinguish between the broadband and periodic components. If
the periods present are not of primary interest, then it is also used 
to evaluate
the overall behavior of the time series. A power law dependence, and thus a
linear dependence on a log-log plot, of $S(f)$ given by

\begin{equation} S(f) ~\sim ~ f^{-\beta} \end{equation} that follows from 
the squared amplitude of the Fourier transform of the signal

\begin{equation} S (f) = \lim _{T\longrightarrow \infty}
\frac{1}{T}\left|\int_{-T}^{T} ~e^{2\pi ift} ~y(t) dt\right|^2,
\end{equation} is a hallmark of a {\it self-affine} phenomenon that 
is underlying
the data. The properties of the signal can be further classified as 
persistent or
anti-persistent fluctuations, depending on the values of the spectral exponent
$\beta$ in Eq. (1). If the signal possesses a tendency for repeating 
the sign of
its fluctuations and, therefore, being of persistent type, then 
$2<\beta<3$; when
fluctuations having opposite signs follow one another, i.e. when 
$1<\beta<2$, the
signal is said to be anti-persistent.\cite{falconer,Turcottebook} A 
signal with a
spectral exponent obeying $1<\beta<3$ is a nonstationary signal with stationary
increments. In this case, the time series created by the increments of a
nonstationary signal has a spectral exponent $\beta$ within the 
interval [-1,1],
and so is a stationary signal.\cite{Turcottebook} The latter property 
facilitates
application of techniques of analysis to the increment signal. However the FT
method, which produces second-order statistics, is insufficient to describe in
full the signal scaling properties, because higher order moments may not be
negligible. Strictly speaking, power spectrum analysis is suitable only for
stationary time series. For more information on spectral time series 
analysis see
[\onlinecite{MalamudTurcotte,Taqqu1995,BA}]. Note that spectral 
analysis provides
information on the scaling properties of the signal at
high-frequency/short-time-scales, in contrast to the detrended fluctuation
analysis method, which is reviewed next.

\subsection{DFA method and time dependence of the correlations}

A method that relaxes the requirement of stationarity of the 
investigated signal
is the detrended fluctuation analysis (DFA) method.\cite{DNADFA} The DFA method
is a tool used for sorting out {\it long range correlations} in a nonstationary
self-affine time series with stationary increments.\cite{nvma,ijmpc,hu} The
method has been used previously in the meteorological field.
\cite{kimaeeta,KoscielnyBundeetal1998,buda} It provides a simple quantitative
parameter - the scaling exponent $\alpha$, which is a signature of the
correlation properties of the signal. Let the signal $y(t)$ be defined between
the beginning $t_0$ and end of the observations $t_M$, i.e. in $[t_0,t_M]$. The
DFA technique consists in dividing a time series $y(t)$ of length $N$ into an
integer number $N/\mu$ of nonoverlapping boxes (called also windows), each
containing $\mu$ points \cite{DNADFA,buda} ($\mu$ = 4,5, ..., ). The 
local trend
$z(n)$ in each box is defined to be the ordinate of a linear 
least-squares fit of
the data points in that box. The detrended fluctuation function $F^2(\mu)$ is
then calculated using:

\begin{equation} F^2(\mu) = {1 \over \mu } {\sum_{n=k\mu+1}^{(k+1)\mu}
{\left[y(n)- z(n)\right]}^2} \qquad \mbox{for} \qquad
k=0,1,2,\dots,\left(\frac{N}{\mu}-1\right). \end{equation}

Averaging $F^2(\mu)$ over the $N/\mu$ intervals gives the mean-square
fluctuations that are assumed to follow a power law

\begin{equation} <F^2(\mu)>^{1/2} \sim \mu^{\alpha}. \label{dfa} \end{equation}

The DFA exponent $\alpha$ so obtained represents the correlation properties of
the signal: $\alpha=1/2$ indicates that the changes in the values of a time
series are random and, therefore, uncorrelated with each other, as in 
a Brownian
random walk sequence. If $\alpha<1/2$ then the signal is anti-persistent
(anti-correlated), and $\alpha>1/2$ indicate positive persistency (correlation)
in the signal. It has been shown by Heneghan and 
McDarby\cite{heneghan} that the
relationship $\beta=1+2\alpha$ holds true for stochastic processes, i.e. for
fractional Brownian walks.

The $\alpha$-exponent value that holds true for a certain time interval called
{\it the scaling range}, is a characteristic of the correlations in the
fluctuations of a signal $y(t)$. It is of interest, however, to test 
whether the
correlations maintain the same properties in shorter intervals within 
$[t_0,t_M]$
or whether they change with time, as it should be anticipated for nonstationary
time series data. In order to probe the existence of so-called {\it locally
correlated} and {\it decorrelated} sequences,\cite{nvma} one can construct an
``observation box'' with a certain width, $\nu$, place the box at the beginning
of the data, calculate $\alpha$ for the data in that box, move the box by
$\Delta\nu$ toward the right along the signal sequence, calculate $\alpha$ in
that box, and so on through the $N$th point of the available data. Each
$\alpha$-value is assigned to the end point of the box because all 
points in the
box are needed in order to obtain $\alpha$. A time-dependent $\alpha$ may be
expected for $t$ ranging from $\nu$ to $N$. This approach is suitable for cloud
base height data because it can be expected to reveal changes in the 
correlation
dynamics of the clouds at various times for a given time lag $\nu$. If such a
time dependence is found, then a multifractal approach is to be
suggested.\cite{nvmamf,barav} In fact, it is expected that the usual fractal
dimension\cite{falconer} $D$ measuring the roughness of a signal\cite{chemnitz}
is directly related to $\alpha$ through\cite{MalamudTurcotte,chemnitz}
$D=2-\alpha$, whence $D$ is time-dependent as well. 

\subsection{Multifractal aspects}

The scaling behavior of a signal $y(t)$ can change in a nonlinear fashion with
the statistical moments, i.e. can be characterized by different scaling
exponents. Such a signal is multi-affine and can be described through
multifractal measures.\cite{davis,711,665,kiandta,canessa} One approach which we
will follow here consists in studying the intermittency\footnote{The notion of intermittency has no canonical
definition\cite{BergePomeauVidal} and covers a variety of phenomena. For us,
intermittency consists in the existence of large and rare 
fluctuations with some
structure $localized$ in space and time for which a peculiar temporal 
behavior is
found between periodic and chaotic regimes.}
and the roughness\cite{711} of the signal. The intermittency is quantified 
adopting the singular measure analysis
The first step that this 
technique requires
is to define a basic measure $\varepsilon(1;l)$ as

\begin{equation} \varepsilon(1;l)=\frac{|\Delta y(1;l)|}{<\Delta 
y(1;l)>}, \qquad
l=0,1, \dots, N -1 \end{equation} where $\Delta 
y(1;l)=y(t_{i+1})-y(t_i)$ is the
small-scale gradient field and $< >$ denotes an average over the $N$ 
data points

\begin{equation}<\Delta y(1;l)>= \frac{1}{N}\sum_{l=0}^{N-1}|\Delta y(1;l)|.
\label{ave} \end{equation}

Next we define a series of ever more coarse-grained and ever shorter fields
$\varepsilon(r;l)$ where $0<l<N-r$ and $r=1,2,4,8\dots$. Thus the 
average measure
in the interval $[l;l+r]$ is

\begin{equation}\varepsilon(r;l)=\frac{1}{r}\sum_{l'=l}^{l+r-1}
\varepsilon(1;l'). \end{equation} The scaling properties of the generating
function are then obtained for through the equation

\begin{equation} <\varepsilon(r;l)^q>\sim \left(\frac{r}{N}\right)^{-K(q)} ,
\quad q\ge 0.
\label{kq}
\end{equation} 
Using $K(q)$-functions one can define\cite{davis} the generalized multifractal 
dimension\cite{grass,proc}

\begin{equation} D(q)=1-\frac{K(q)}{q-1}. 
\label{dkq}
\end{equation} 
Note that at $q=1$ the l'Hospital rule is used to obtain D(1).

Further, the multi-affine properties of a time-dependent signal $y(t)$ can also be
described by the so-called $q$th order structure functions\cite{davis}

\begin{equation} \left<|y(t_{i+r}) - y(t_i)|^q \right> \sim 
\tau^{qH(q)}, \qquad
i=1,2, \dots , N -r \end{equation} with $\tau \equiv \Delta_r=t_{i+r}-t_i$.
Whence $H(q)$ and $K(q)$ describe the multifractal scaling properties of
nonlinear dynamic processes. In a monofractal case, $H(q)$ and and $K(q)$ take
constant values.

\subsection{Hierarchy of exponents $h(\gamma)$}

The multi-affinity of $y(t)$ means that one should use different scaling
exponents $H(q)$ in order to rescale a signal in various scaling 
ranges. In other
words, this also implies that {\it local} scaling exponent $\gamma$ 
exists\cite{nvmamf} in order to characterize the local singularity of
the signal. The density of points $N_{\gamma}(\Delta)$
that have the same local scaling exponent is usually assumed\cite{mf} to scale over
the time span $\Delta$ (for any $r$) as \begin{equation}
N_{\gamma}(\Delta)\sim \Delta^{-h(\gamma)}. \end{equation}

From Ref. [\onlinecite{bara}] the following relations are found:

\begin{equation} \gamma(q)=\frac{d(qH(q))}{dq} \label{gq} \end{equation}

\begin{equation} h(\gamma(q))=1+q\gamma(q)-qH(q) \label{hg}\end{equation} The
$h(\gamma(q))$ function\cite{nvmamf} is as naturally adapted to describe
multi-affine signals as the multifractal spectrum for multifractal
objects.\cite{mf}

Several of the methods reviewed in this section have been recently used in the
meteorological
field.\cite{kimaeeta,KoscielnyBundeetal1998,buda,davis,kiandta,Tsonisetal1,Tsonisetal2,TalknerWeber,JAM}

\section{Data and data analysis}

Next we apply the above methods to study the nonstationarity of cloud base
height (CBH) data sets and the correlations in their fluctuations {\it on all
measured time scales}; data are obtained at the Southern Great Plains 
site of the
Atmospheric Radiation Measurement Program of the Department of Energy using a
Belfort Laser Ceilometer (BLC) Model 7013C.\cite{arm} The ceilometer is a
self-contained, ground-based, optical, active, remote sensing 
instrument with the
ability to detect and process several cloud-related parameters, among 
them cloud
base height, cloud extinction coefficient and cloud layer depth. The ceilometer
system detects clouds by transmitting pulses of infrared light vertically into
the atmosphere and analyzing the returned signals backscattered by the
atmosphere. The receiver telescope detects scattered light from clouds and
precipitation.\cite{ceilometer} The ceilometer actively collects backscattered
photons for about 5 seconds within every 30-second measurement period. The BLC
measures the base height of the lowest cloud detected between 15 and 7350 m
directly above mean ground level.

The cloud base height signal measured on Sept. 23-25, 1997 is plotted in Fig.1.
Data consists of $N=7251$ data points, as for the purpose of this analysis we
consider the record from 5:15 UTC on Sept. 23, through 17:40 UTC on Sept. 25,
1997. It is well representative of similar events often occurring on a shorter
time interval.

The probability density function (PDF) for the cloud base height 
signal (data in
Fig.1) is shown in Fig.2. The values of the abscissa, $\Delta y=y(t+\Delta
t)-y(t)$ are unnormalized and therefore, given in meters. The cases $\Delta
t=30$~s and $\Delta t=30$~min are displayed in Fig.2. They correspond to short
and long time lags. Both PDFs are strongly non-Gaussian. The double pyramid
triangular shape with a width growing with increased time lag is 
similar to that
found in other meteorological cases.\cite{genpol,varna}

We first check the stationarity of the data by applying a power spectral
analysis. The power spectral density of the cloud base height data set is shown
in Fig. 3. The upper curve represents the spectral density that is smoothed by
applying a moving average over equidistant intervals on the log-scale. This
procedure is anticipated to be better suited to identifying scaling 
properties of
the spectrum.\cite{FT} It should be noted that the procedure may lead 
to a slight
change in the slope of the linear fit that may to alter the scaling exponent
slightly, as seen here from $\beta=1.46\pm0.08$ to $\beta=1.33\pm0.06$ for
frequencies lower than 1/15 min$^{-1}$, i.e. $\sim$ 10$^{-3}$ Hz. Nevertheless,
because $1<\beta<3$, we can conclude that, the cloud base height data are
nonstationary with stationary increments. Futhermore, because $1<\beta<2$, the
signal is anti-persistent. Note that the non-smoothed spectrum is more suitable
for detecting characteristic frequencies and periods in the raw 
signal. The error
bars in this paper are calculated according to standard
techniques.\cite{statbook}

Similar values for the spectral exponent, e.g. $\beta$=1.28$\pm$0.1 and
$\beta$=1.49$\pm$0.08, are obtained in Ref. 
[\onlinecite{genpol,varna}] for cloud
base height data measurements during the Atmospheric Stratocumulus Transition
Experiment (ASTEX). ASTEX was designed to clarify the transition from
stratocumulus to trade cumulus clouds\cite{garratt} in the marine 
boundary layer
in the region of the Azores Islands.\cite{mbl} Studies of more cases are needed
to clarify whether the slight difference in the scaling exponents 
between stratus
over land, which is the subject of this paper, and stratocumulus clouds over
ocean is physically and statistically significant or not.

The DFA leads to an $\alpha$ exponent equal to $0.18\pm0.002$ for 
time lags less
than 55~min, followed by an exponent $0.13\pm0.001$, as indicated in 
the inset of
Fig.4. Notice that the Heneghan and McDarby\cite{heneghan} relationship
$\beta=1+2\alpha$ holds well for the shorter time lag region. Incidently we
stress that two spectral regimes are hardly seen from the above spectral power
density graph in Fig.3.

The time dependence of the correlations is next discussed. Results 
are plotted in
Fig. 4 as a function of the cloud life time for $\nu=7$~h and $\Delta
\nu=30$~min. The $\nu$ value is chosen such that the finite-size effects are
avoided, while the $\Delta \nu$ value is somewhat arbitrarily chosen for an
adequate display. Smaller values of $\Delta\nu$ lead to rougher curves
representing the time-dependence of $\alpha$, nevertheless without 
significantly
altering the overall results.

The generalized fractal dimensions $D(q)$ are shown in Fig. 5 and seen to
decrease with $q$. The multi-affinity of the CBH signal is also 
observed from the
nonlinearity of the functions $qH(q)$, $\gamma(q)$ and $K(q)$ in Figs.5 and 6.
The value of the $H(q)$-function at $q=1$ defines the nonstationarity parameter
$H_1$ that is a measure of the roughness of the signal, while $C_1$ at $q=1$

\begin{equation} C_1 = \left .\frac{dK(q)}{dq}\right|_{q=1} \label{c1}
\end{equation} usually defines the degree of intermittency of the
signal.\cite{davis,nvmamf,711,665}

The value obtained for the $H(q)$-function at $q=1$, 
$H(q=1)=H_1=0.21\pm0.02$ is close
to the $\alpha$-exponent of DFA method, $\alpha=0.18\pm0.002$ of this study and
is similar to the results for the cloud base height data measured during ASTEX,
for which\cite{genpol} $\alpha=0.24\pm0.002$ for June 18, 1992 and
$\alpha=0.21\pm0.005$ for June 15, 1992 and for which\cite{varna} 
$H(q=1)=H_1=0.23\pm0.04$ for June 14, 1992 and
$H(q=1)=H_1=0.21\pm0.03$ for June 15, 1992. If the cloud base height data series
was a strictly defined monofractal signal, then one can expect that $\alpha=H_1$.\cite{falconer}

 Note that $q=1$ is a special case that corresponds to a monofractal
behavior. Using l'Hospital's rule in Eq. (\ref{c1}) one can obtain the
information dimension of the CBH signal, if it scales as a monofractal
\begin{equation} D(1) = 1 - \left.\frac{dK(q)}{dq}\right|_{q=1} \end{equation}
The dashed line in Fig. 5 defines the monofractal case 
$D(1)=1-C_1=0.90$. Having
in mind that $C_1$ is related to the mean of $\varepsilon(r;l)$ for a
one-dimensional field, we note that the singularities that contribute 
most to the
$<\varepsilon(r;l)>$ occur on a set with fractal dimension $D(1)$. In order to
compensate statistically for their sparse spatial distribution, in 
sharp contrast
to a Gaussian process, these extreme events are far more rare but far more
intense. There is no intermittency at all if $D(q)\equiv 1$, i.e. $C_1=0$.

The multi-affine properties of the signal are represented by the $h(\gamma)$
functions following Eq. (\ref{hg}) in Fig.7. Although the $h(\gamma)$-curve does
not reach unity, it tends to some maximum value that 
corresponds to $\gamma_0
\sim 0.21$, a value equal to the $\alpha$ exponent of the signal 
(Fig. 4). The crossing of the $x$-axis by the $h(\gamma)$-curve defines the
minimum value $\gamma_{min}\sim 0$ that is related to the minimum value of the $\alpha$
exponent that is contained in the signal.\cite{nvmamf} Here the value of
$h(0)=0.70$, in agreement with the value 1-$qH(q)$ at its maximum as obtained
from Fig.6, (see Eqs. (\ref{gq}) and (\ref{hg})).

\section{Conclusions}

It is often asked what the numerical values of such above exponents mean, and
whether they have any physical interpretation at all. Let us recall that the
fractal dimension of a profile is a measure of its roughness, while generalized
fractal dimensions have been given some thermodynamic interpretation
elsewhere\cite{domany,kadabook} and here above (Sect.2). The above $\alpha$
exponent indicates that the cloud bottom is  quite rough. The occurrence of two
scaling regimes might be understood either from a mathematical point 
of view or a
physical one. In the former case, Hu et al.\cite{hu} have indicated 
that such an
occurrence can be due to different noises and different trends. On the other
hand, the presence of two scaling regimes well separated by a 
crossover time lag
may indicate that two processes occur in stabilizing, or not, the cloud base
height data set (or profile). Indeed we had observed that the $\alpha$-exponent
tends toward low values when the cloud is in quasi-equilibrium and 
reaches a high
value when the cloud breaks apart\cite{kimaeeta}. We may conjecture 
here that the
two scaling exponents describe two regimes of long time span (slow) 
correlations,
which maintain stable droplets, and one at shorter time scales for fast
correlations between droplets for both agglomeration or cloud fracture process.
More cases need to be studied, however, in order to  provide  a more precise
interpretation of these so close values of the $\alpha$-exponents.

Let us also recall that the simplest interpretation of a stochastic signal is
through the notion of (fractional or not) Brownian motion, or random 
walk. In so
doing, one can interpret the CBH signal as mimicking the distance 
from the origin
(the initial time) traveled by a particle diffusing, e.g. on a lattice. An
anomalous diffusion occurs if the signal correlations are not strictly
characterized by the laws governing classical Brownian motion, but by 
other types
of power laws.\cite{BouchaudGeorges,MetzlerKlafter} If  those power laws are
characterized by time-dependent exponents, then multifractality is expected.
Usually the fractal dimension of a strange attractor measures the 
minimum number
of components of the phase space necessary to describe the dynamic process. No
such fractal dimension has been studied here. It would be interesting 
to look for
it in order to define how many interrelated physical parameters are 
necessary to
describe the CBH variability, i.e. pressure, temperature, humidity, wind
velocity, etc. Following that line of thought, we can consider the CBH as
representing stochastically driven  turbulent eddies, in which phase 
transitions
occur in three-dimensional space. In so doing the CBH is exactly a visual
observation of these transitions in eddies  moving up and down and horizontally
in a stochastic way, as a random walk. In our case, the data set 
being studied is
not really representing a particle in a 3D space but
rather a projection on a 1D space; only the bottom of the cloud,
  projected on a vertical axis is
studied with respect to its up and down motion, - just like particle undergoing
(fractional or not) Brownian motion in a plane.

In summary, after recalling a few statistical analysis techniques, we have
applied them to a typical continental stratus cloud base height 
profile data set.
We have demonstrated that the CBH profile is a nonstationary anti-persistent
signal with stationary increments. The spectral exponent found has a value
similar to the one for stratoculumus clouds over the ocean reported 
by some of us
in another study. The multi-affine scaling properties of the data series found
reflect the complexity of the processes that produce them. Further 
work should be
directed toward relating the scaling properties expressed through these
statistical parameters to the dynamical properties of the clouds, an important
step toward understanding, modeling and predicting their dynamical behavior.

\section{Acknowledgments}

This research was partially supported by Battelle grant number 327421-A-N4. We
acknowledge collaboration of the U.S. Department of Energy as part of the
Atmospheric Radiation Measurement Program. MA thanks A. P\c{e}kalski 
for comments.

 \newpage

{\large \bf Figure Captions} \vskip 0.5cm

{\noindent \bf Figure 1} -- Cloud base height signal measured on Sept. 23-25,
1997 at the Southern Great Plains site of the Atmospheric Radiation Measurement
Program of the Department of Energy. The abscissa marks the time in 
hours after 0
UTC on Sept. 23, 1997. The data series contains 7251 data points.

\vskip 0.5cm {\noindent \bf Figure 2} -- Probability density function $P(\Delta
y,\Delta t)$ (PDF) of the cloud base height signal measured on Sept. 
23-25, 1997
(data in Fig. 1). The values on the abscissa, $\Delta y$, are given in meters.
Triangles correspond to short time lag $\Delta t=30$~s equal to the
discretization step of the data series, and circles mark the long time lag PDF
with $\Delta t=30$~min. None of the PDF curves is displaced.

\vskip 0.5cm {\noindent \bf Figure 3} -- Power spectrum $S(f)$ of the 
cloud base
height signal measured on Sept. 23-25, 1997. The upper curve represents the
smoothed spectra, vertically displaced by two decades, that scales with a
spectral exponent $\beta=1.33\pm0.06$.

\vskip 0.5cm {\noindent \bf Figure 4} -- The local $\alpha$-exponent (black
circles) given by the Detrended Fluctuation Analysis method as a 
function of time
for the cloud base height (CBH) data measured on Sept. 23-25, 1997. Error bars
are indicated. The rescaled (divided by a factor 3000) CBH signal is shown by
dots. Inset: Log-log plot of the DFA-function Eq.(\ref{dfa}), showing 
the scaling
exponents $\alpha_1=0.18$, $\alpha_2=0.13$, and the crossover time 
lag at 55 min.

\vskip 0.5cm {\noindent \bf Figure 5} -- Hierarchy of generalized dimensions
$D(q)$ for the CBH data in Fig.1. The straight line is drawn to 
enhance the value
$q_s$ at which the $D(q)$ function starts to deviate from a linear dependence.
The dashed line defines the monofractal case $D(1)=0.90$; from 
Eq.(\ref{c1}) one
thus finds $C_1$=0.10. The $K(q)$-function defined through Eq. (\ref{kq})
indicates the intermittency of the signal.

\vskip 0.5cm {\noindent \bf Figure 6} -- The hierarchy of exponents $qH(q)$ and
$\gamma(q)$ indicating the multi-affine properties of the continental stratus
cloud base height (CBH) data measured on Sept. 23-25, 1997. For $q=1$,
$H_1=0.21\pm0.02$ is a parameter of nonstationarity.

\vskip 0.5cm {\noindent \bf Figure 7} -- The $h(\gamma)$-curve for the stratus
continental cloud base height data measured on Sept. 23-25, 1997; 
data in Fig. 1.

\newpage \begin{figure}[ht] \begin{center} \leavevmode \epsfysize=8cm
\epsffile{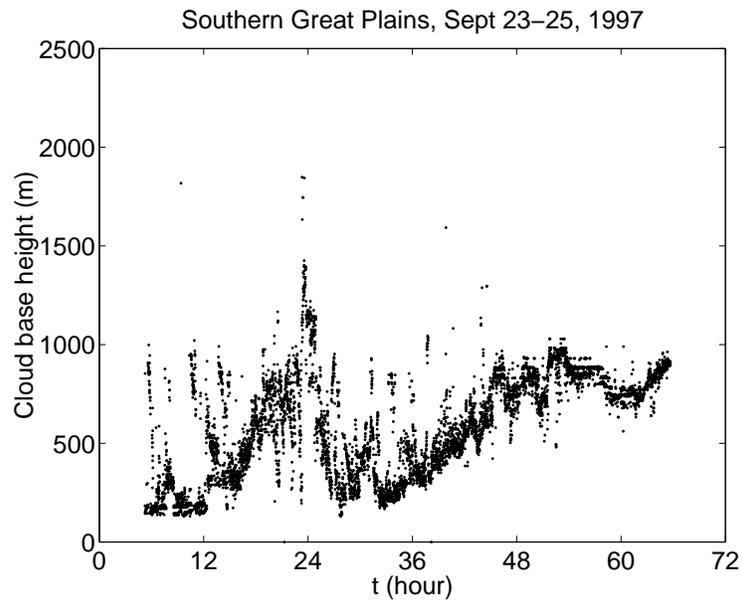} \caption{Cloud base height signal measured on Sept. 23-25,
1997 at the Southern Great Plains site of the Atmospheric Radiation Measurement
Program of the Department of Energy. The abscissa marks the time in 
hours after 0
UTC on Sept. 23, 1997. The data series contains 7251 data points.} \end{center}
\end{figure}

\newpage \begin{figure}[ht] \begin{center} \leavevmode \epsfysize=8cm
\epsffile{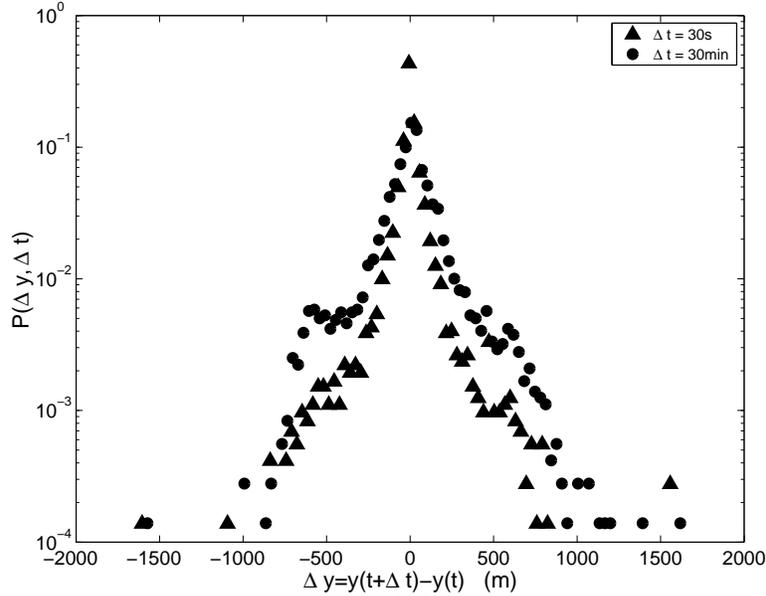} \caption{Probability density function $P(\Delta 
y,\Delta t)$
(PDF) of the cloud base height signal measured on Sept. 23-25, 1997 
(data in Fig.
1). The values on the abscissa, $\Delta y$, are given in meters. Triangles
correspond to short time lag $\Delta t=30$~s equal to the 
discretization step of
the data series, and circles mark the long time lag PDF with $\Delta 
t=30$~min. None of the PDF curves is displaced.}
\end{center} \end{figure}

\begin{figure}[ht] \begin{center} \leavevmode \epsfysize=8cm 
\epsffile{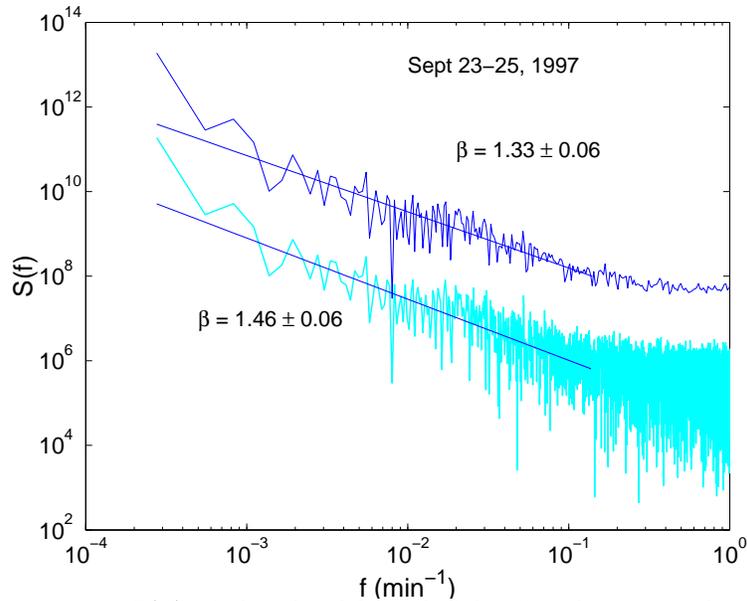}
\caption{Power spectrum $S(f)$ of the cloud base height signal 
measured on Sept.
23-25, 1997. The upper curve represents the smoothed spectra, vertically
displaced by two decades, that scales with a spectral exponent
$\beta=1.33\pm0.06$.} \end{center} \end{figure}

\begin{figure}[ht] \begin{center} \leavevmode \epsfysize=8cm 
\epsffile{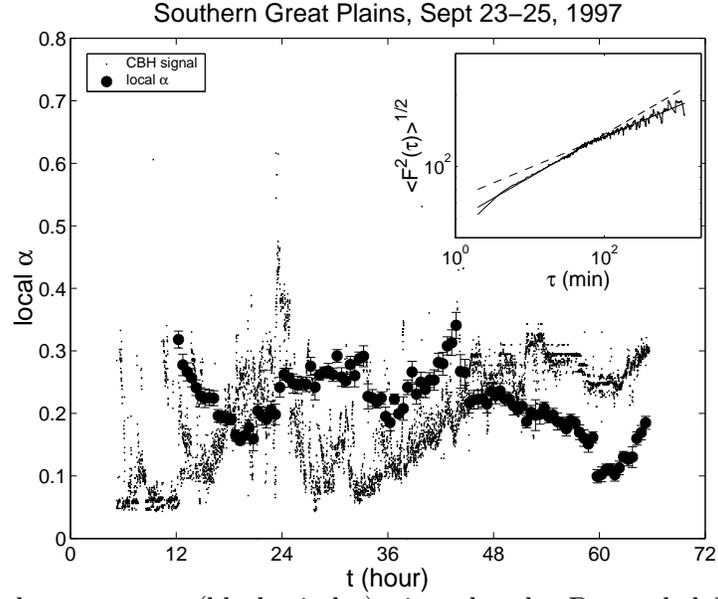}
\caption{The local $\alpha$-exponent (black circles) given by the Detrended
Fluctuation Analysis method as a function of time for the cloud base 
height (CBH)
data measured on Sept. 23-25, 1997. Error bars are indicated. The rescaled
(divided by a factor 3000) CBH signal is shown by dots. Inset: Log-log plot of
the DFA-function Eq.(\ref{dfa}), showing the scaling exponents $\alpha_1=0.18$,
$\alpha_2=0.13$, and the crossover time lag at 55 min.} \end{center} 
\end{figure}

\newpage \begin{figure}[ht] \begin{center} \leavevmode \epsfysize=8cm
\epsffile{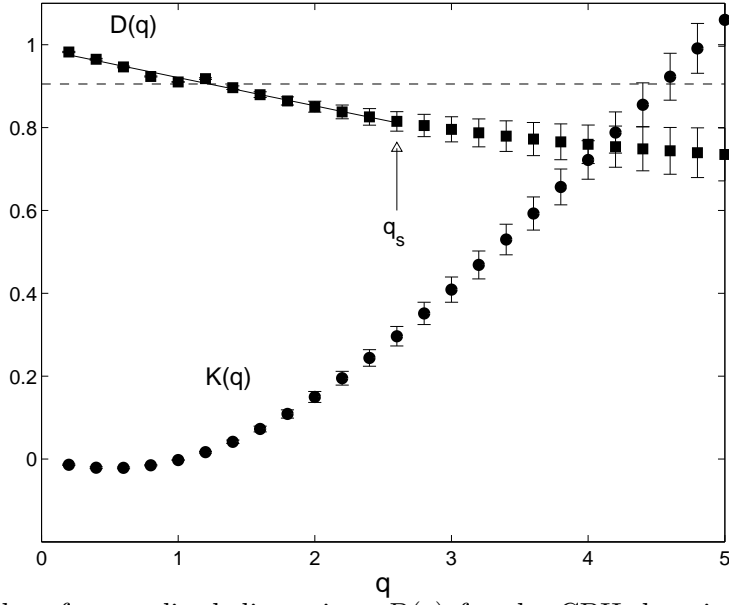} \caption{Hierarchy of generalized dimensions 
$D(q)$ for the
CBH data in Fig.1. The straight line is drawn to enhance the value 
$q_s$ at which
the $D(q)$ function starts to deviate from a linear dependence. The dashed line
defines the monofractal case $D(1)=0.90$; from Eq.(\ref{c1}) one thus finds
$C_1$=0.10. The $K(q)$-function defined through Eq. (\ref{kq}) indicates the
intermittency of the signal.} \end{center} \end{figure}

\begin{figure}[ht] \begin{center} \leavevmode \epsfysize=8cm 
\epsffile{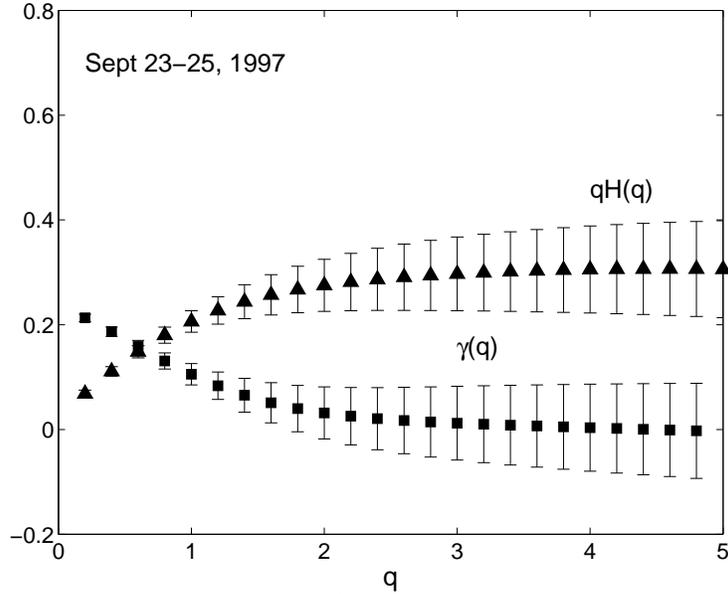}
\caption{The hierarchy of exponents $qH(q)$ and $\gamma(q)$ indicating the
multi-affine properties of the continental stratus cloud base height (CBH) data
measured on Sept. 23-25, 1997. For $q=1$, $H_1=0.21\pm0.02$ is a parameter of
nonstationarity.} \end{center} \end{figure}

\begin{figure}[ht] \begin{center} \leavevmode \epsfysize=8cm 
\epsffile{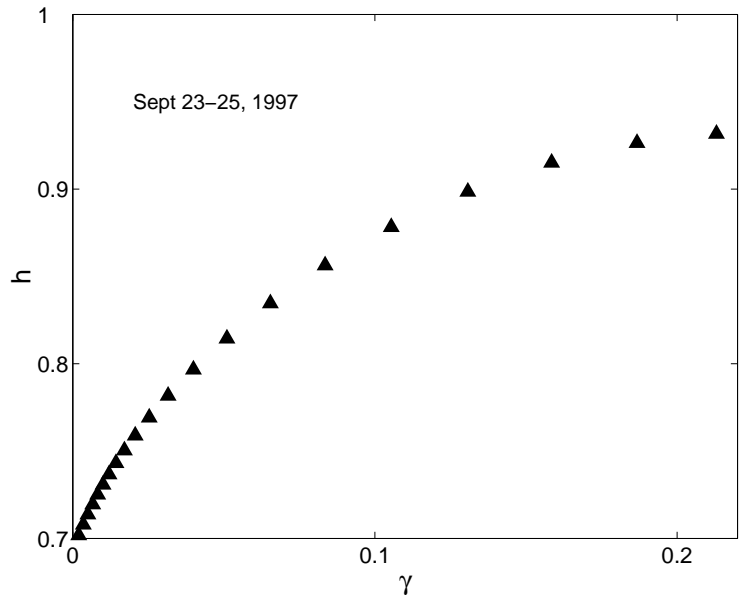}
\caption{The $h(\gamma)$-curve for cloud base height data measured on Sept.
23-25, 1997.} \end{center} \end{figure}

\end{document}